# FIFA Does it Right: 2026 FIFA World Cup Does not Increase the Number of Non-Competitive Matches

Traian Marius Truta

Computer Science Department, Northern Kentucky University,
Highland Heights, KY 41076, U.S.A.
trutat1@nku.edu

**Abstract.** FIFA World Cup is one of the most watched sporting event in the world and its popularity continue to increase. While currently 32 teams participate in this event, starting with 2026 the number of participants will increase to 48. As a result, FIFA proposed a major format change, the groups of 4 teams will be replaced by groups of 3 teams, with the first two teams to qualify to the second round. In this paper, we discuss the possibility of matches with limited importance (called non-competitive) for final group standing for both 4-team and 3-team groups and we introduce our definitions of collusion and dead rubber games. Next, we discuss scenarios in both 4-team and 3-team group formats that lead to such matches. Finally, we make an analysis of all group stage matches from last six World Cup editions (1998 to 2018), and we predict the most likely number of collusions and dead rubber games for the new format of the World Cup. During our analysis we consider a variation of the proposed 3-group format in which the order of matches for each group is predetermined. Our analysis show that this variation has the potential of reducing significantly the number of both collusion and dead rubber matches thus becoming an interesting option to be considered for adoption by FIFA. In conclusion, we recommend FIFA to implement a predefined order of group matches prior to the drawing. The seeded team should play the first match with the team that is second best (according to the drawing time FIFA Coca-Cola rankings), as well as the second match from the group.

## 1  Introduction and Motivation

The Fédération Internationale de Football Association (FIFA) is an international governing body of association football, futsal, and beach soccer [3]. FIFA is responsible for the organization of football's major international tournaments, notably the World Cup. FIFA World Cup is one of the most watched sporting event in the world and its popularity continue to increase. This event started in 1930, with its first edition hosted in Uruguay. Since then the World Cup had brought together the best national soccer teams every four years (except 1942 and 1946 due to World War II) [4]. This year, the 21 edition of the FIFA World Cup was hosted by Russia [7].

In this prestigious sporting event, national teams, preselected after very competitive qualifying matches, compete for the FIFA World Cup Trophy. Due to increasing popularity of the sport, increase demand for more matches, and the desire for more profits, the number of teams that participate in the World Cup had increase overtime. Initially only up to 16 teams participated in the World Cup (up to 1978, three editions - 1930, 1934, and 1950 – had fewer than 16 participants), then the number of teams was increased to 24 (1982 – 1994), then to 32 (1998 – 2022) and finally the number of teams will increase to 48 starting with 2026 World Cup. In general and for all more recent editions (1954 and forward) the World Cup started with groups of four teams. For instance for the most recent editions (1998 – 2018, and this format will likely be used in 2022) the 32 qualified teams are divided in 8 groups of 4 teams. The first two from each group qualifies for the second knock-out round. In this second round there are 8 round of 16 matched followed by 4 quarterfinals, two semifinal the 3-place final, and the final. The total number of matches is 64 (48 in the first round and 16 in the second round) [15].

For 2026 edition, in addition to increasing the number of participants to 48, FIFA proposed a major format change, the groups of 4 teams will be replaced by groups of 3 teams. As a result there will be 16 groups in the first round with a total of 48 matches. The first two from each group will then play in the second round. The second round will be similar to the current format with the addition of the Round of 32 matches. One major reason of changing the well-established format of this event was the constraint of not increasing the number of matches for individual teams. Indeed, the first 4 teams will play 7 matches in both formats, teams ranked 5 – 8 will play 5 matches, teams ranked 9 – 16 will play 4 matches, teams ranked 17 – 32 will play 3 matches in both formats. In addition, in the new format, teams ranked 33 to 48 will play 2 matches each and they will also be part of the World Cup event [1].

While it initially seems that there are only advantages in the new format, the groups of 3 teams have the potential of reducing the quality of some matches which are between teams of different levels [12]. Other issues have been reported, in particular the danger of increased number of collusion games [9].

In this paper we make the following contributions. First, we present our definitions of collusion and dead rubber games (we group both of them under the non-competitive game umbrella) and we describe scenarios in either 4-group or 3-group format that lead to such matches. Second, we look at all 3-group games in FIFA World cup history and discuss if they had non-competitive games. Third, and our major contribution, based on the results of the last six World Cup editions (1998 to 2018), we compare the number of collusion and dead rubber matches from these editions and we predict the number of collusions and dead rubber games for the new format of the World Cup.

The main motivation before starting the data analysis for this paper was to analyze if the new format of the World Cup will significantly increase the number of non-competitive matches due to 3-team groups. At this point, it is not too late for FIFA to make changes to the format (such as considering the 4-teams groups in the first round) that will increase the attractiveness of the tournament.

## 2 Non Competitive Matches Scenarios and Definitions

### 2.1 Notations

We assume that there are $n$ teams in a group. The teams from this group are labeled $T_i$, where $i = 1\ldots n$. We also assume that $T_i$ is ranked higher in the FIFA Coca-Cola ranking then $T_j$ for all $i \leq j$. The FIFA Coca-Cola ranking is a ranking system for men's national soccer teams based on all FIFA recognized international matches [5]. We assume that the first k teams ($k < n$) qualify for the second round. We call such a group a ($n$, $k$)-group.

The ranking of teams is determined by the rules used currently in 2018 FIFA World Cup [6]:
1. Greatest number of points obtained in all group matches (3 points per win, 1 per draw, 0 per loss);
2. Goal difference in all group matches;
3. Greatest number of goals scored in all group matches.
4. If two or more teams are equal on the basis of Steps 1 – 3, their ranking will be determined by applying to the group matches between the equal teams the following rules:
    a. Greatest number of points obtained in the group matches between the teams concerned;
    b. Goal difference resulting from the group matches between the teams concerned;
    c. Greater number of goals scored in all group matches between the teams concerned;
    d. Greater number of points obtained in the fair play conduct of the teams based on yellow and red cards received in all group matches;
    e. Drawing of lots by the FIFA.

### 2.2 Analysis of Non Competitive Matches from 2018 FIFA World Cup

Based on above notations, in the current format (1998 – 2022 editions), there were 8 (4, 2)-groups in the first round. Starting from 2026, the intent is to have 16 (3, 2)-groups. Also, please note that the last two matches from a (4, 2)-group start at the same time, and the below analysis takes this factor into consideration.

In each group, the most interesting matches is when each team will have an incentive to obtain a minimum result in order to qualify for the next round. Below, we analyses the possible scenarios before the last two matches in a (4, 2)-group. We present examples based on FIFA World Cup 2018 groups.

A. A team, $T_i$, is already qualified before a match (last match for small values of n such as 3 or 4). This team may not use its best line-up for that match and the match will be of reduced interest to viewers. That being said, the final ranking in a group is also important since the team ranked higher plays with a team ranked lower in the different group. So the team may still have a limited incentive. The second team from the match, $T_j$, may be in the following scenarios (there is no order relation between *i* and *j*, in other words $T_i$ may be ranked either lower or higher in the FIFA Coca Cola rankings):

1. The second team in this match may already qualified too. In that case there is only a limited incentive (the ranking order). This scenario represents a ***dead rubber*** game. For the 2018 World Cup we had the following games in this category:
   - Russia vs Uruguay (0 – 3).
   - England vs Belgium (0 – 1).

   For the first match, Russia performed poorly since they did not have high hopes of qualifying before the start of the tournament and the results shows this fact. For the second match both teams played mostly with reserves [10] and some news articles even hinted that England wanted to lose this game in order to avoid being in the same quarter with Brazil [11].

2. The second team in this match will need a minimum result to qualify. This scenario represents a ***collusion*** game. For the 2018 World Cup we had the following game in this category:
   - France vs Denmark (0 – 0).

   In this game, France was already qualified and Denmark needed a draw to qualify regardless of the result of the other match (Peru vs Australia). Although Denmark had a small incentive of wining the group, they were happy with a draw result and this match was one of the least interesting matches of the entire World Cup [2].

   A second match almost fits this category:
   - Croatia vs Iceland (2 – 1).

   In this match, Croatia was already qualified and Iceland could qualify with win depending of the results from the second match (Argentina vs Nigeria). Croatia still had the incentive to draw in order to secure the first place in the ranking. This game also represents a ***collusion*** game, however the collusion did not happen.

3. The second team in this match may be already eliminated. In that case there is only a limited incentive (the ranking order) for the qualified team and the desire to perform well in their last match for the team already eliminated. This scenario is impossible in both (4, 2) and (3, 2)-groups. That being said, there are situations that fit this scenario unless a very strange result will happen. Assume a (3, 2) group with the following 2 results: $T_1$ vs $T_2$ (4 – 0), $T_2$ vs $T_3$ (5 – 0). In this case, $T_2$ is already qualified and $T_1$ will lose the qualification only if will lose by 5 goals of more in the match against $T_3$. Since this is very unlikely ($T_1$ is considered the best team in the group), we call this situation a ***weak dead rubber***. Assume now that the first two results in the group were: $T_1$ vs $T_2$ (4 – 0), $T_2$ vs $T_3$ (4 – 0). In this case the $T_1$ and $T_2$ are almost qualified, however the result $T_1$ vs $T_3$ (1 – 5), will eliminate $T_2$ and qualify $T_1$ and $T_3$. In this case, the match becomes a ***collusion***, although could also be considered as a weak dead rubber.

B. A team, $T_i$, is already eliminated before a match (last for small values of n such as 3 or 4). The second team from the match, $T_j$, may be in the following scenarios

(there is no order relation between $i$ and $j$, in other words $T_i$ may be ranked either lower or higher in the FIFA Coca Cola rankings):
1. The second team in this match may be qualified. This is similar to A.3 case.
2. The second team in match will need a minimum result to qualify. This scenario represents a ***collusion*** game. For the 2018 World Cup we had the following matches in this category:
   - Spain vs Morocco (2 – 2).
   - Germany vs Korea (0 – 2).
   - Switzerland – Costa Rica (2 – 2).
   - Poland vs Japan (1 – 0).

   Only the first and the third match ended in the collusion results, and both matches were interesting to watch. The other two matches did not have collusion results. It is worth noting that in Poland vs Japan due to the result of the simultaneous match (Senegal vs. Columbia) the final result was still a collusion result and the last 10 minutes of the game showed that none of the teams wanted to do anything. Also, Germany vs Korea was one of the most interesting matches from the World Cup since it lead to the elimination of the 2014 World Cup winners, Germany.
3. The second team in this match may be already eliminated. This scenario represents a ***dead rubber*** game. For the 2018 World Cup we had the following matches in this category:
   - Egypt vs Saudi Arabia (1 – 2)
   - Tunisia vs Panama (1 – 0)

   In addition the following game was a likely dead rubber game since a draw in the other match would eliminate both teams.
   - Peru vs Australia (2 – 0)

   We call this match a ***weak dead rubber*** since a likely result from the other group match that starts at the same time will seal the fate for both teams.

Overall, in 2018 FIFA World Cup, four matches were dead rubbers, one was weak dead rubber, and six collusions out of which three results were collusion results.

## 2.3 Definitions

Based on the analysis performed in the previous subsection, we define the following categories of non-competitive matches during a group stage at a FIFA World Cup event.

***Collusion (C)*** - There is at least one result between $T_a$ and $T_b$ that maximizes $T_a$ chances to qualify in the next round to the detriment to one other team ($T_c$). The result qualifies team $T_b$ as well, or the status of team $T_b$ is already known before this game ($T_b$ is either qualified or eliminated prior to this match).
This definition includes all six collusions matches presented above.

***Met Collusion (MC)*** – A collusion match became a met collusion if the final result satisfy the collusion requirement.
Three of the sex collusion matches from 2018 World Cup were met collusions.

***Aggravated Collusion (AC)*** - A collusion between $T_a$ and $T_b$ where one team may lose its last match by any score and it still finishes first in the group.

None of the above six collusions fits this criteria, so at least the final ranking was still in play for all of the collusion matches. However, for a (4, 2) – group, there is a case when aggravated collusion could happen. Assume $T_a$ wins against $T_b$ and $T_c$ and $T_d$ draws with $T_b$ and $T_c$. The final match between $T_a$ and $T_d$ is an aggravated collusion since $T_a$ has already 6 points and cannot lose first place, and $T_d$ will qualify with a win. Such a scenario does not happen often, the most recent ones were in 1998 FIFA World Cup (2 such aggravated collusions): Brazil vs Norway (1 – 2) and Paraguay – Nigeria (3 – 1). Brazil had six points, Norway had two points before the final match. The final result met the aggravated collusion and as a result Morocco was eliminated. While there were no suspicions of match fixing, still the results show that since Brazil did not have any incentive to win they had a poor match. Similarly, the second result was a met aggravated collusion with Nigeria being in the same situation as Brazil and Paraguay in the same situation as Norway.

Please note that we use a modified definition for *AC* compared to [8] since a team may try to win by multiple goals to improve its ranking.

***Weak Collusion (WC)*** – There is at least one result between $T_a$ and $T_b$ that allows $T_a$ to qualify in the next round to the detriment to one other team ($T_c$), unless $T_c$ scores at least $g$ goals in its last match. We use in this paper $g$ equal to 3. The result qualifies team $T_b$ as well, or the status of team $T_b$ is already known before this game ($T_b$ is either qualified or eliminated prior to this match).

Note that a weak collusion is not a collusion game. There were no weak collusions in the 2018 World Cup. We illustrate a weak collusion later in this section when we discuss the scenarios for a (3, 2)-group.

For both aggravated and weak collusions, we also have the corresponding **met aggravated collusion (MAC)** and **met weak collusion (MWC)** matches. Note that the above mentioned Brazil vs Norway and Paraguay vs Nigeria matches were *MAC* matches.

***Dead Rubber (D)*** - A match where regardless of the score both teams are either qualified or eliminated.

This definition includes all four dead rubber matches presented in the beginning of this section.

***Weak Dead Rubber (WD)*** - A match where regardless of the score may matter only if in a different match the score is not a collusion score or if a team needs to win by $g$ or more goals. We use again $g$ equal to 3.

This definition includes Peru vs Australia match from 2018 World cup edition discussed in this section. As in the case of collusions, a weak dead rubber is not a dead rubber game.

There is no equivalent of met collusion for dead rubber games. However, we could define **Aggravated Dead Rubber (AD)** and **Aggravated Weak Dead Rubber (AWD)** matches when the ranking order between the two team does not change regardless of the score. None of those two types of matches are possible for either (4, 2) and (3, 2)-groups.

## 2.4 Analysis of Non Competitive Matches for a (3, 2)-group

We assume that $T_a$, $T_b$, $T_c$ are playing the matches in the following order: $T_a$ vs $T_b$, $T_a$ vs $T_c$, and $T_b$ vs $T_c$. In this analysis we do not use the ranking of teams (i. e. we do not know which team is ranked higher before the tournament). Note that the team that plays two matches will qualify regardless of the last match result (between the other two teams) if it has at least three points and a positive goal difference. Also, a team is eliminated if it has zero points after two matches. See [8] for a complete proof of these facts.

A. $T_a$ wins by 3 goals or more in the first match. In this case. The second match is a *WC* match, since a $T_b$ win by a smaller margin (2 goals or even 1 in most cases) will lead to the requirement that $T_c$ needs a win by at least 3 goals in the last match to qualify. This is the only scenario when the second match fits into a non-competitive category. If this *WC* game became a *MWC* (the result has been met), then the third match became a *WD* match.

B. $T_a$ gets at least 4 points in the first two matches. Then no collusion or dead rubber matches can happen.

C. $T_a$ gets 3 points in the first two matches and has a positive goal difference. The last match is likely a competitive match unless the team that has 0 points needs a win by 3 goals or more. In that case the last match can be a *WD* match.

D. $T_a$ gets 3 points in the first two matches and has a negative or zero goal difference. The last match became a *C* match since there is a result that will qualify both $T_b$ and $T_c$, and eliminate $T_a$. Note that such a result is usually not very likely.

E. $T_a$ gets 2 points in the first two matches. The last match became a *C* match since there is a result (high scoring draw) that will qualify both $T_b$ and $T_c$, and eliminate $T_a$. Note that such a result is not very likely.

F. $T_a$ gets 1 points in the first two matches. The last match became a *C* match since there are many results that will qualify both $T_b$ and $T_c$, and eliminate $T_a$. Converting this *C* match into a *MC* match is very likely (any draw met the collusion results).

G. $T_a$ gets 0 points in the first two matches. The last match became a *D* match both $T_b$ and $T_c$, are already qualified.

Based on the above analysis, the desired outcome for a group with respect to maintain all matches competitive is that the team that plays first will qualify after the first two matches without winning with a large goal difference in the first match. The probability that this will happen in a group increases when $T_a$ is the higher ranked team prior to the start of the tournament, $T_b$ is the second ranked team in the group (this will limit the high difference scores in the first round), and $T_c$ is the weaker team. In Section 3, we analyze based on real results the impact that such scheduling will have in limiting the non-competitive matches. Also, note that aggravated collusion (as well as aggravated (weak) dead rubbers) are not possible in a (3, 2)-group format.

## 2.5 Three Group Teams in the History of the World Cup

There have been several World Cups when groups of 3 teams have been used. These situations are presented below.

In 1930, due to lack of participants, 3 groups were formed by 3 teams only. Only the first team qualified, so this groups were (3, 1) – groups. Two of the groups had the last match as a dead rubber match. The format was changed for the next edition.

In 1950, one group had 3 participants but one withdrew (France) and the group became an eliminatory match.

In 1982, (3, 1) – groups were used in the second round. To avoid dead rubber matches, the team that won the first match played the third match. This solved the non-competitive matches' problem, but created logistics issues since the teams did not know their schedule for the second and third match in advance. This format was also dropped after 1982 edition.

Since 1982, FIFA did not use a 3 team group in its flagship competition. However a (3, 2) – group was never tested and 2026 will be the first World Cup where such a format will be used.

## 3 Non-competitive Matches Prediction for 2026 FIFA World Cup

To predict how many matches will be non-competitive in the 2026 FIFA World Cup, we analyze the results from the group round for the past six World Cup editions (1998 to 2018). The reason of choosing the last six editions is that they all use the same format of qualification thus the analysis between all the editions is comparable. We will compute the number of *C*, *WC*, *D*, and *WD* in each World Cup, we will also use these results to simulate all possible (3, 2)-groups. In each (3, 2) – group we use the results in order of which they were played as well as in the order discussed in Section 2 (best team per FIFA Coca-Cola Ranking plays the first two matches, and the first match is against the second best team).

### 3.1 2018 FIFA World Cup Analysis

We present the 2018 FIFA World Cup analysis in greater detail below. All the results presented below are used from the official 2018 FIFA World Cup site [4]. We consider Group A for a complete discussion.

In Fig.1, we show the standing before the last two matches in the group. Since the last two matches were played at the same time, the collusions / dead rubbers must be determined at this moment. The columns have the following meaning: **Pos** – position in the group, **Team** – the name of the team, **Pld** – number of matches played by the Team, **W** – number of matches won, **D** – number of matches that are draw, **L** – number of matches that are lost, **GF** – goals for this team (number of goals scored by Team), **GA** – goals against (number of goals received by the team), **GD** – goal differential, and **Pts** – number of points. All these columns are standard columns in a soccer classification. The last columns, **Rank**, has the following meaning – the first number represents the rank between the teams in the group at the drawing time based on FIFA Coca-Cola ranking [5], the second number represents the actual ranking in the FIFA Coca-Cola ranking. Please note that the host nation (Russia) will automatically receive the order 1 regardless of the actual FIFA ranking. Note that FIFA Coca-Cola Ranking History is fully available at [5]. In Fig. 2, we show the game results. The last two matches (in red) were played at the same time after the standings are as shown in Fig. 1.

| Pos | Team | Pld | W | D | L | GF | GA | GD | Pts | Rank |
|-----|------|-----|---|---|---|----|----|----|-----|------|
| 1 | 🇷🇺 Russia (H) | 2 | 2 | 0 | 0 | 8 | 1 | +7 | 6 | 1 - 65 |
| 2 | 🇺🇾 Uruguay | 2 | 2 | 0 | 0 | 2 | 0 | +2 | 6 | 2 - 21 |
| 3 | 🇪🇬 Egypt | 2 | 0 | 0 | 2 | 1 | 4 | −3 | 0 | 3 - 31 |
| 4 | 🇸🇦 Saudi Arabia | 2 | 0 | 0 | 2 | 0 | 6 | −6 | 0 | 4 - 63 |

**Fig. 1.** 2018 FIFA World Cup – Group **A** Standings before last two matches.

| Russia | 5 - 0 | Saudi Arabia |
|--------|-------|--------------|
| Uruguay | 1 - 0 | Egypt |
| Russia | 3 - 1 | Egypt |
| Uruguay | 1 - 0 | Saudi Arabia |
| Russia | 0 - 3 | Uruguay |
| Egypt | 1 - 2 | Saudi Arabia |

**Fig. 2.** 2018 FIFA World Cup – Group **A** Results.

As already discussed in Section 2, we can easily determine that Russia vs Uruguay (both already qualified) and Egypt vs Saudi Arabia (both eliminated) are *D* matches.

For each group there are 4 possible combinations of 3 teams that may represent a possible group starting with 2026 FIFA World Cup. We label these groups derived from Group A with A1, A2, A3, and A4. In group A1 we have the teams ranked 1, 2, and 3 at the draw (in this case Russia, Uruguay, Egypt); in group A2, teams ranked 1, 2, and 4; in group A3, teams ranked 1, 3, and 4, and in group A4, teams ranked 2, 3, and 4. For each group we consider the following two order of games. First, the order is the same as the actual order. We call this order Real Order. These groups will be finally labelled as A1a, A2a, A3a, and A4a. Second, the order is of games is as follows, team ranked 1 plays first with team ranked 2. Second match is between team ranked 1 and team ranked 3, and finally the teams 2 and 3 will play the last match. We call this order 1-2-3. These groups will be finally labelled as A1b, A2b, A3b, and A4b. These 8 groups are shown in Fig. 3 with standings after the second result. Note that for A4a and A4b we list an additional column (Y – representing the number of yellow cards) which due to the perfect tie in points and goals determines the ranking between Egypt and Saudi Arabia. For each of those 8 possible groups via direct observation we determine any collusion or dead rubber matches.

| Pos | Team | Pld | W | D | L | GF | GA | GD | Pts | *C/WC/D/WD* Matches |
|---|---|---|---|---|---|---|---|---|---|---|
| 1 | Russia | 1 | 1 | 0 | 0 | 3 | 1 | +2 | 3 | *D*: RUS – URU |
| 2 | Uruguay | 1 | 1 | 0 | 0 | 1 | 0 | +1 | 3 | |
| 3 | Egypt | 2 | 0 | 0 | 2 | 1 | 4 | −3 | 0 | |
| | | | | | | | | | | **Group A1a** |

| Pos | Team | Pld | W | D | L | GF | GA | GD | Pts | *C/WC/D/WD* Matches |
|---|---|---|---|---|---|---|---|---|---|---|
| 1 | Uruguay | 1 | 1 | 0 | 0 | 3 | 0 | +3 | 3 | *C*: URU – EGY (0 – 2) |
| 2 | Russia | 2 | 1 | 0 | 1 | 3 | 4 | -1 | 3 | Not a *MC* |
| 3 | Egypt | 1 | 0 | 0 | 1 | 1 | 3 | −2 | 0 | |
| | | | | | | | | | | **Group A1b** |

| Pos | Team | Pld | W | D | L | GF | GA | GD | Pts | *C/WC/D/WD* Matches |
|---|---|---|---|---|---|---|---|---|---|---|
| 1 | Russia | 1 | 1 | 0 | 0 | 5 | 3 | +2 | 3 | *D*: RUS – URU |
| 2 | Uruguay | 1 | 1 | 0 | 0 | 1 | 0 | +1 | 3 | |
| 3 | Saudi Arabia | 2 | 0 | 0 | 2 | 0 | 6 | −6 | 0 | |
| | | | | | | | | | | **Group A2a** |

| Pos | Team | Pld | W | D | L | GF | GA | GD | Pts | *C/WC/D/WD* Matches |
|---|---|---|---|---|---|---|---|---|---|---|
| 1 | Uruguay | 1 | 1 | 0 | 0 | 3 | 0 | +3 | 3 | *WD*: URU – KSA |
| 2 | Russia | 2 | 1 | 0 | 1 | 5 | 3 | +2 | 3 | (SA needs to win by 4+ goals; not happened) |
| 3 | Saudi Arabia | 1 | 0 | 0 | 1 | 0 | 5 | −5 | 0 | |
| | | | | | | | | | | **Group A2b** |

| Pos | Team | Pld | W | D | L | GF | GA | GD | Pts | *C/WC/D/WD* Matches |
|---|---|---|---|---|---|---|---|---|---|---|
| 1 | Russia | 2 | 2 | 0 | 0 | 8 | 1 | +7 | 6 | *WC*: RUS – EGY (0 – 1) (2nd game) |
| 2 | Egypt | 1 | 0 | 0 | 1 | 1 | 3 | −2 | 0 | Not a *MWC* |
| 3 | Saudi Arabia | 1 | 0 | 0 | 1 | 0 | 5 | −5 | 0 | |
| | | | | | | | | | | **Group A3a** |

| Pos | Team | Pld | W | D | L | GF | GA | GD | Pts | *C/WC/D/WD* Matches |
|---|---|---|---|---|---|---|---|---|---|---|
| 1 | Russia | 2 | 2 | 0 | 0 | 8 | 1 | +7 | 6 | None |
| 2 | Egypt | 1 | 0 | 0 | 1 | 1 | 3 | −2 | 0 | |
| 3 | Saudi Arabia | 1 | 0 | 0 | 1 | 0 | 5 | −5 | 0 | |
| | | | | | | | | | | **Group A3b** |

| Pos | Team | Pld | W | D | L | GF | GA | GD | Pts | Y | *C/WC/D/WD* Matches |
|---|---|---|---|---|---|---|---|---|---|---|---|
| 1 | Uruguay | 2 | 2 | 0 | 0 | 2 | 0 | +2 | 6 | - | None |
| 2 | Saudi Arabia | 1 | 0 | 0 | 2 | 0 | 1 | −1 | 0 | 0 | |
| 3 | Egypt | 1 | 0 | 0 | 2 | 0 | 1 | −1 | 0 | 2 | |
| | | | | | | | | | | | **Group A4a** |

| Pos | Team | Pld | W | D | L | GF | GA | GD | Pts | Y | *C/WC/D/WD* Matches |
|---|---|---|---|---|---|---|---|---|---|---|---|
| 1 | Uruguay | 2 | 2 | 0 | 0 | 2 | 0 | +2 | 6 | - | None |
| 2 | Saudi Arabia | 2 | 0 | 0 | 2 | 0 | 6 | −1 | 0 | 0 | |
| 3 | Egypt | 2 | 0 | 0 | 2 | 1 | 4 | −1 | 0 | 2 | |
| | | | | | | | | | | | **Group A4b** |

**Fig. 3.** Groups **A1a, A1b, A2a, A2b, A3a, A3b, A4a,** and **A4b** before last match.

In Fig 3, in addition to the team 3 letter code [14] we include the following information for each type of non-competitive game. For a D match – nothing; for a WD match – why is a weak dead rubber, how many goals one team needs to win by; for a C match – a possible collusion result, or set of results, if the collusion is a met collusion or not; and for a WC match a possible collision result and if the collusion was net or not. Note that in this group possible scenarios we have all types of non-competitive matches. Also, in this figure we label 6 out of 8 (3, 2)-groups with red (A1a, A1b, A2a, A2b, A3a, and A3b). The meaning of this color is that such a group is more likely to exist in 2026 tournament based on FIFA Coca-Cola ranking at drawing time of participating teams. We consider a group to be a possible group if one of the team is ranked 1-15 or host country (in this case Russia), and the other 2 teams are ranked 16 or more. We assume that the group drawing procedure is similar to the 2018 drawing and 16 seeds will be preselected based on their ranking.

In Fig 4, we summarize all the non-competitive matches for the original (4, 2)-group A, as well as all 8 combinations of possible (3, 2)-group. The color red has the same meaning of highlighting a possible group based on team ranking and drawing procedure. For a collusion match, *no* means that the collusion result did not happen, while *yes* means that the match was a met collusion. We use the color blue to show the (4, 2)-group.

| Group | C | WC | D | WD |
|---|---|---|---|---|
| A | | | RUS – URU<br>EGY – KSA | |
| A1a | | | RUS – URU | |
| A2a | | | RUS – URU | |
| A3a | | RUS – EGY (0-1), *no* | | |
| A4a | | | | |
| A1b | URU – EGY (0-2), *no* | | | |
| A2b | | | | URU – KSA, KSA 4+ win |
| A3b | | | | |
| A4b | | | | |

**Fig. 4.** 2018 FIFA World Cup - Group **A** analysis of non-competitive matches.

Using the above specified approach we are able to perform the same analysis for all remaining 7 groups of the World Cup. Since all the results are available [4], we will just illustrate the final analysis for each group in Appendix A (we include for completeness Group A as well).

The summary of the non-competitive matches for all 8 groups as well as for their corresponding (3, 2)-groups is shown in Fig. 5. The last four columns have the following meaning: *TC* – total collusions (*C* + *WC*); *TD* – total dead rubbers (*D* + *WD*); *TC+TD* – total collusions and dead rubbers (non-competitive matches); and *MTC* – met total collusions. Since the number of matches for (3, 2)-groups is higher than for a real tournament (where the number of matches is 48), we weight all the numbers to a total of 48 matches. This new summary which allow an easier comparisons between (4, 2)-groups and (3, 2)-groups is presented in Fig. 6.

| 2018 World Cup | Total Matches | C | WC | D | WD | TC | TD | TC+TD | MTC |
|---|---|---|---|---|---|---|---|---|---|
| (4, 2) - groups | 48 | 6 | 0 | 4 | 1 | 6 | 5 | 11 | 3 |
| (3, 2) – groups; real order | 96 | 6 | 1 | 10 | 0 | 7 | 10 | 17 | 2 |
| (3, 2) – groups; 1-2-3 order | 96 | 9 | 0 | 1 | 1 | 9 | 2 | 11 | 2 |
| (3, 2) – groups; real order; red groups | 51 | 5 | 1 | 4 | 0 | 6 | 4 | 10 | 2 |
| (3, 2) – groups; 1-2-3 order; red groups | 51 | 7 | 0 | 0 | 1 | 7 | 1 | 8 | 2 |

**Fig. 5.** 2018 FIFA World Cup - Non-competitive matches summary.

| 2018 World Cup | C | WC | D | WD | TC | TD | TC+TD | MTC |
|---|---|---|---|---|---|---|---|---|
| (4, 2) - groups | 6 | 0 | 4 | 1 | 6 | 5 | 11 | 3 |
| (3, 2) – groups; real order | 3 | 0.5 | 5 | 0 | 3.5 | 5 | 8.5 | 1 |
| (3, 2) – groups; 1-2-3 order | 4.5 | 0 | 0.5 | 0.5 | 4.5 | 1 | 5.5 | 1 |
| (3, 2) – groups; real order; red groups | 4.71 | 0.94 | 3.76 | 0.00 | 5.65 | 3.76 | 9.41 | 1.88 |
| (3, 2) – groups; 1-2-3 order; red groups | 6.59 | 0.00 | 0.00 | 0.94 | 6.59 | 0.94 | 7.53 | 1.88 |

**Fig. 6.** 2018 FIFA World Cup - Non-competitive matches summary weighted to 48 matches (2026 FIFA World Cup format).

### 3.2 Previous Five Editions Analysis

We performed the same analysis for the previous five editions of the FIFA World Cup. We include in this subsection only non-competitive non-weighted matches' summaries (Fig. 7-11. Per group analysis as well as both summaries (non-weighted and weighted) are presented in Appendices B (2014 World Cup), C (2010 World Cup), D (2006 World Cup), E (2002 World Cup), and E (1998 World Cup). All the results are available in the FIFA World Cup Archive [7] and, therefore, they are not included in this paper.

| 2014 World Cup | Total Matches | C | WC | D | WD | TC | TD | TC+TD | MTC |
|---|---|---|---|---|---|---|---|---|---|
| (4, 2) - groups | 48 | 4 | 0 | 3 | 1 | 4 | 4 | 8 | 1 |
| (3, 2) – groups; real order | 96 | 13 | 0 | 5 | 1 | 13 | 6 | 19 | 6 |
| (3, 2) – groups; 1-2-3 order | 96 | 6 | 0 | 2 | 1 | 6 | 3 | 9 | 0 |
| (3, 2) – groups; real order; red groups | 39 | 6 | 0 | 2 | 0 | 6 | 2 | 8 | 2 |
| (3, 2) – groups; 1-2-3 order; red groups | 39 | 2 | 0 | 0 | 1 | 2 | 1 | 3 | 0 |

**Fig. 7.** 2014 FIFA World Cup - Non-competitive matches summary.

| 2010 World Cup | Total Matches | C | WC | D | WD | TC | TD | TC+TD | MTC |
|---|---|---|---|---|---|---|---|---|---|
| (4, 2) - groups | 48 | 2 | 0 | 1 | 2 | 2 | 3 | 5 | 1 |
| (3, 2) – groups; real order | 96 | 15 | 0 | 4 | 1 | 15 | 5 | 20 | 7 |
| (3, 2) – groups; 1-2-3 order | 96 | 9 | 0 | 2 | 1 | 9 | 3 | 12 | 3 |
| (3, 2) – groups; real order; red groups | 51 | 8 | 0 | 2 | 1 | 8 | 3 | 11 | 4 |
| (3, 2) – groups; 1-2-3 order; red groups | 51 | 5 | 0 | 1 | 1 | 5 | 2 | 7 | 2 |

**Fig. 8.** 2010 FIFA World Cup - Non-competitive matches summary.

| 2006 World Cup | Total Matches | C | WC | D | WD | TC | TD | TC+TD | MTC |
|---|---|---|---|---|---|---|---|---|---|
| (4, 2) - groups | 48 | 6 | 0 | 4 | 3 | 6 | 7 | 13 | 3 |
| (3, 2) – groups; real order | 96 | 10 | 1 | 8 | 0 | 11 | 8 | 19 | 5 |
| (3, 2) – groups; 1-2-3 order | 96 | 6 | 0 | 4 | 0 | 6 | 4 | 10 | 2 |
| (3, 2) – groups; real order; red groups | 51 | 4 | 1 | 2 | 0 | 5 | 2 | 7 | 2 |
| (3, 2) – groups; 1-2-3 order; red groups | 51 | 2 | 0 | 0 | 0 | 2 | 0 | 2 | 1 |

**Fig. 9.** 2006 FIFA World Cup - Non-competitive matches summary.

| 2002 World Cup | Total Matches | C | WC | D | WD | TC | TD | TC+TD | MTC |
|---|---|---|---|---|---|---|---|---|---|
| (4, 2) - groups | 48 | 4 | 0 | 0 | 2 | 4 | 2 | 6 | 1 |
| (3, 2) – groups; real order | 96 | 10 | 1 | 5 | 1 | 11 | 6 | 17 | 4 |
| (3, 2) – groups; 1-2-3 order | 96 | 6 | 1 | 1 | 1 | 7 | 2 | 9 | 4 |
| (3, 2) – groups; real order; red groups | 60 | 6 | 1 | 3 | 1 | 7 | 4 | 11 | 2 |
| (3, 2) – groups; 1-2-3 order; red groups | 60 | 3 | 1 | 1 | 1 | 4 | 2 | 6 | 2 |

**Fig. 10.** 2002 FIFA World Cup - Non-competitive matches summary.

| 1998 World Cup | Total Matches | C | WC | D | WD | TC | TD | TC+TD | MTC |
|---|---|---|---|---|---|---|---|---|---|
| (4, 2) - groups | 48 | 4 | 1 | 3 | 4 | 5 | 7 | 12 | 4 |
| (3, 2) – groups; real order | 96 | 13 | 0 | 5 | 0 | 13 | 5 | 18 | 6 |
| (3, 2) – groups; 1-2-3 order | 96 | 4 | 2 | 4 | 0 | 6 | 4 | 10 | 2 |
| (3, 2) – groups; real order; red groups | 39 | 3 | 0 | 1 | 0 | 3 | 1 | 4 | 1 |
| (3, 2) – groups; 1-2-3 order; red groups | 39 | 0 | 2 | 0 | 1 | 2 | 1 | 3 | 1 |

**Fig. 11.** 1998 FIFA World Cup - Non-competitive matches summary.

### 3.2 Results

In Figs. 12 – 16, we compare the number of non-competitive matches between the analyzed World Cup editions. The five bars for each World Cup represent (from left to right), the number of matches in that category from the (4, 2)-groups (the actual groups of the World Cups), the number of matches form (3, 2)-groups when the order of play is the original order, the number of matches from (3, 2)-groups when the order of play is 1-2-3, the number of matches form (3, 2)-groups when the order of play is the original order and the group is a more realistic group (red color; exactly one team ranked in the first 16 teams), and the number of matches from (3, 2)-groups when the order of play is 1-2-3 and the group is a more realistic group. The *avg* represents the average between all editions and it represents our predicted values for 2026 World Cup.

Fig. 12 show that the total number of non-competitive matches will likely decrease in the (3, 2)-format. The decrease is small if the order of play is random, however the 1-2-3 order of play makes a significant improvement reducing by almost half the number of non-competitive matches.

Fig. 13 show that the number of collusion matches will have a slight increase for a random order, however it will decrease for a 1-2-3 order. We do not distinguish between regular collusion and weak collusion, however the number of weak collusions is significantly lower.

Fig. 14 show the major improvement, the number of dead matches will significantly decrease for all (3, 2)-group scenarios. Still the biggest decrease is for 1-2-3 order for the realistic group scenarios. We do not distinguish between regular dead rubbers and weak dead rubbers, however the number of weak dead rubbers is significantly lower.

Figs. 15 and 16 show the met collusions results. It is interesting to see that the percentage of met collusion is higher for (4, 2)-groups and this is because when the teams know that a collusion score is possible they do not have a significant incentive to achieve a different result. Still even in this case the percentage is below 50% and this leads to the conclusion that such matches are still of interest.

To make a prediction/guess of the number of noncompetitive matches we average the results obtained for (3, 2)-groups scenarios (all groups and red groups) for the real order of games and we round them to the nearest integer. We will get the following likely values:

- Total Collusions = 6
- Total Dead Rubbers = 3
- Total Non-Competitive Matches = 9
- Total Met Collusion = 3 (the average is 2 however due to attractiveness of collusion score, we increase this value by 1).

Using a similar approach we also predict the number of non-competitive matches for 1-2-3 order. We will get the following likely values:

- Total Collusions = 3
- Total Dead Rubbers = 1
- Total Non-Competitive Matches = 4
- Total Met Collusion = 1

Overall, the proposed format will be similar to the current format in terms of non-competitive matches, there will be more collusion matches than dead rubber matches. While this can be seen as a problem, from the game appeal to spectators this is still an improvement since the dead rubber matches have the lowest TV audience ratings from comparable matches. For instance England vs Belgium and Uruguay vs Russia (both dead rubber matches) had the lowest TV ratings in US between all final group matches broadcasted by FOX. Also Panama vs Tunisia and Saudi Arabia vs Egypt (also dead rubber matches) had the lowest TV ratings in US between all final group matches broadcasted by Fox Sports 1. In general, collusion matches had higher TV ratings, although their ratings also suffers compared to competitive matches [14].

A significant improvement is when 1-2-3 order is used, and in this case the number of non-competitive matches is significantly reduced. Having an average of 4-5 more competitive matches (with two fewer dead rubbers) will increase the attractiveness of the group stage significantly and *we strongly recommend FIFA to implement this relatively minor change in the scheduling format*.

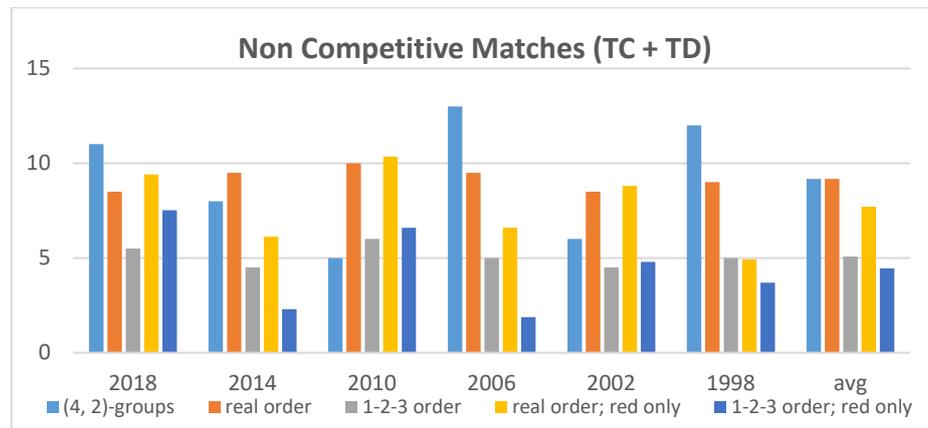

**Fig. 12.** Non-competitive matches comparison.

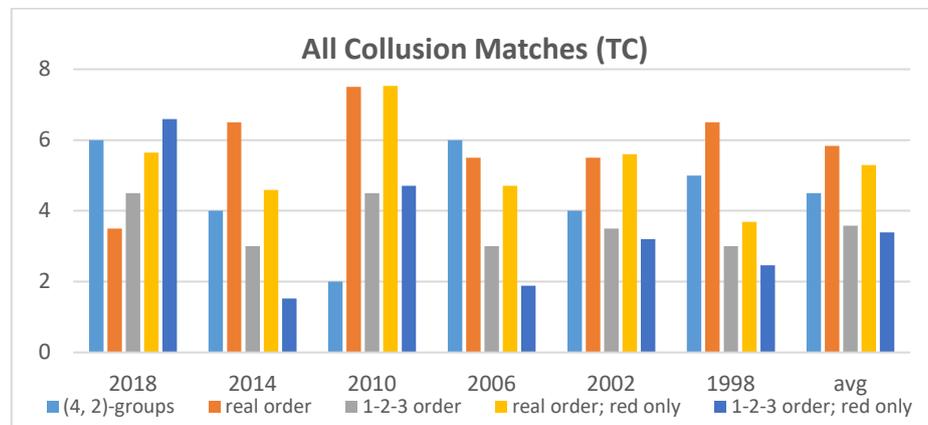

**Fig. 13.** Collusion matches comparison.

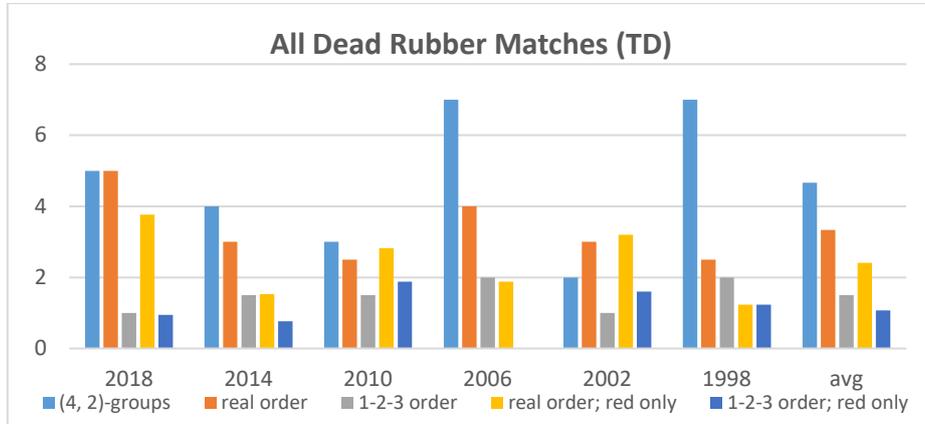

**Fig. 14.** Dead Rubber matches comparison.

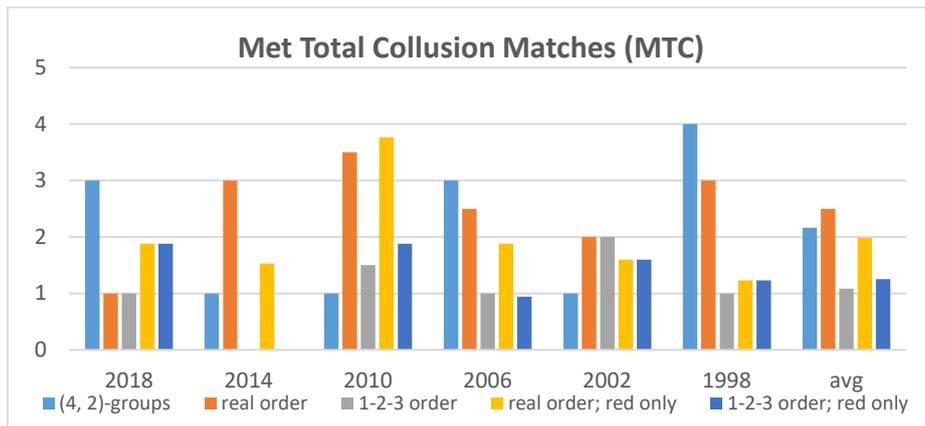

**Fig. 15.** Met collusion matches comparison.

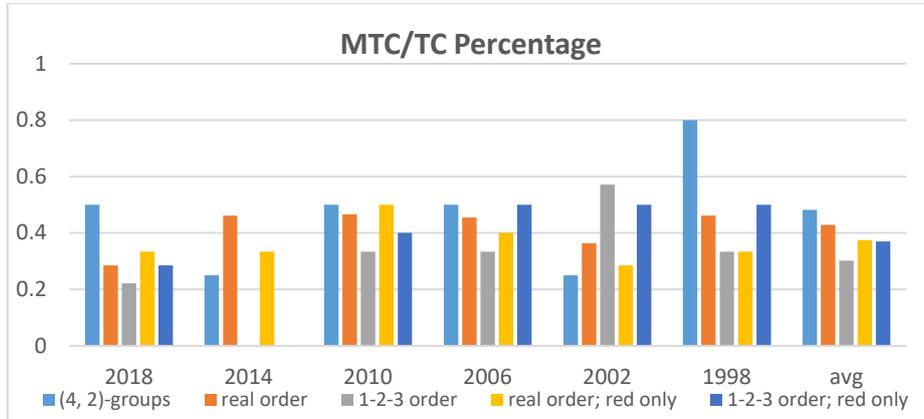

**Fig. 16.** The percentage of met collusion matches from total collusion matches.

## 3 Discussion and Conclusion

The current FIFA World Cup format is highly successful. However there are matches that are of reduced interest due to the previous results. We classified these matches in collusion and dead rubber matches. We provided an extensive analysis of when such matches take place and we looked at previous six World Cup editions in order to predict how many such matches will likely happen in the new FIFA World Cup format that will be used starting with 2026 edition.

Based on our analysis we claim that the new format will likely have the same number of non-competitive matches, however a slight improvement of game scheduling in the group stage will reduce this number by roughly half and therefore will increase the attractiveness of such matches.

*We strongly recommend FIFA to implement a predefined order of group matches. The seeded team should play the first match with the team that is second best (according to the drawing time FIFA Coca-Cola rankings), as well as the second match from the group.*

The recommendation that 3-groups format with predefined game scheduling is better than 3-groups format with random scheduling is also present in [8]. However, their recommendation is based on reducing the number of collusions only where their definition of collusion is more restrictive then in this paper (suspicion of collusion occurs when a result between two teams lets both of them advance at the expense of a third team [8]).

Having this predefined order may seem that will disadvantage the seeded team since if the seeded team do not perform well there is a possibility of collusion between the second and third best teams in the group (this issue was also discussed in [8]). However, this will happen less frequently than will random order of game scheduling. More importantly the seeded team will have more rest days between their last group match and the Round of 32, and this is an important advantage. Any seeded team will aim and hope to qualify directly bay having at least 3 points and a positive goal difference, thus the additional rest days seems to be more important for such a team.

From the second best team point of view, they will start with their most difficult match. They have the rest days between group matches and this may give them and advantage versus the third ranked team. They will know exactly what result they need with that team in order to qualify. Thus, such order is a favorable order for a second best team as well.

Surprisingly the third best team may not be disadvantaged either. If the seeded team wins the first match, they will aim to obtain a better result than the second team and this means that a draw will suffice in the last match. The main disadvantage is that they will start the competition with rest days, however they will be in the best position to know what results they need in order to maximize their qualification chances.

Last but not least, starting the World Cup with better teams will increase the attractiveness of the first matches and will make people that are casual viewers to follow the World Cup from the beginning. What would you prefer, to see England versus Belgium as the first group match when they both need a good result, or as last group match when both are qualified?

# Appendix A –2018 FIFA World Cup per Group Analysis for Non-Competitive Matches

In Appendix A, we show the non-competitive matches that could occur in all possible (3, 2)-groups that are based on actual (4, 2)-groups in the 2018 FIFA World Cup [4]. Information regarding the non-competitive match is included in below images. Empty cells means there were no matches in that category. The matches' summaries are shown in the last two figures.

| Group | C | WC | D | WD |
|---|---|---|---|---|
| A |  |  | RUS – URU<br>EGY – KSA |  |
| A1a |  |  | RUS – URU |  |
| A2a |  |  | RUS – URU |  |
| A3a |  | RUS – EGY (0-1), *no* |  |  |
| A4a |  |  |  |  |
| A1b | URU – EGY (0-2), *no* |  |  |  |
| A2b |  |  |  | URU – KSA, KSA 4+ win |
| A3b |  |  |  |  |
| A4b |  |  |  |  |

**Fig. A.1.** 2018 FIFA World Cup - Group **A** analysis of non-competitive matches.

| Group | C | WC | D | WD |
|---|---|---|---|---|
| B | ESP – MAR (ESP win/draw), **yes (2-2)** |  |  |  |
| B1a |  |  |  |  |
| B2a |  |  |  |  |
| B3a |  |  | IRN – POR |  |
| B4a | ESP – MAR (1-2), no |  |  |  |
| B1b | ESP – IRN (2-2), no |  |  |  |
| B2b |  |  |  |  |
| B3b |  |  |  |  |
| B4b |  |  |  |  |

**Fig. A.2.** 2018 FIFA World Cup - Group **B** analysis of non-competitive matches.

| Group | C | WC | D | WD |
|---|---|---|---|---|
| C | FRA – DEN (0-0), **yes** |  |  | PER – AUS |
| C1a |  |  | FRA – DEN |  |
| C2a |  |  |  |  |
| C3a |  |  | FRA – DEN |  |
| C4a |  |  |  |  |
| C1b |  |  |  |  |
| C2b |  |  |  |  |
| C3b |  |  |  |  |
| C4b |  |  |  |  |

**Fig. A.3.** 2018 FIFA World Cup - Group **C** analysis of non-competitive matches.

| Group | C | WC | D | WD |
|---|---|---|---|---|
| **D** | CRO – ISL (0-1), no | | | |
| **D1a** | CRO – ISL (most), **yes (1-2)** | | | |
| **D2a** | | | | |
| **D3a** | ARG – NGA (most), **yes (1-2)** | | | |
| **D4a** | CRO –ISL (1-3), no | | | |
| **D1b** | CRO – ISL (most), **yes (1-2)** | | | |
| **D2b** | CRO – NGA (0-1), no | | | |
| **D3b** | | | | |
| **D4b** | | | | |

**Fig. A.4.** 2018 FIFA World Cup - Group **D** analysis of non-competitive matches.

| Group | C | WC | D | WD |
|---|---|---|---|---|
| **E** | SUI – CRC (SUI win/draw), **yes (2-2)** | | | |
| **E1a** | | | | |
| **E2a** | | | | |
| **E3a** | | | BRA - SRB | |
| **E4a** | | | | |
| **E1b** | | | | |
| **E2b** | | | | |
| **E3b** | | | | |
| **E4b** | | | | |

**Fig. A.5.** 2018 FIFA World Cup - Group **E** analysis of non-competitive matches.

| Group | C | WC | D | WD |
|---|---|---|---|---|
| **F** | GER – KOR (GER win), no | | | |
| **F1a** | MEX – SWE (2-1), no | | | |
| **F2a** | | | | |
| **F3a** | GER – KOR (1-2), no | | | |
| **F4a** | | | MEX – SWE | |
| **F1b** | MEX – SWE (2-1), no | | | |
| **F2b** | | | | |
| **F3b** | SWE – KOR (1-0), **yes** | | | |
| **F4b** | SWE – KOR (0-1), no | | | |

**Fig. A.6.** 2018 FIFA World Cup - Group **F** analysis of non-competitive matches.

| Group | C | WC | D | WD |
|---|---|---|---|---|
| **G** | | | BEL – ENG<br>TUN – PAN | |
| **G1a** | | | BEL – ENG | |
| **G2a** | | | BEL – ENG | |
| **G3a** | | | | |
| **G4a** | | | | |
| **G1b** | | | | |
| **G2b** | | | | |
| **G3b** | | | | |
| **G4b** | | | | |

**Fig. A.7.** 2018 FIFA World Cup - Group **G** analysis of non-competitive matches.

| Group | *C* | *WC* | *D* | *WD* |
|---|---|---|---|---|
| **H** | POL – JPN (JPN win/draw), no | | | |
| **H1a** | | | COL – SEN | |
| **H2a** | | | | |
| **H3a** | | | | |
| **H4a** | | | | |
| **H1b** | | | COL – SEN | |
| **H2b** | | | | |
| **H3b** | SEN – JPN (2 – 3), no | | | |
| **H4b** | SEN – JPN (3 – 2), no | | | |

**Fig. A.8.** 2018 FIFA World Cup - Group **H** analysis of non-competitive matches.

| 2018 World Cup | *Total Matches* | *C* | *WC* | *D* | *WD* | *TC* | *TD* | *TC+TD* | *MTC* |
|---|---|---|---|---|---|---|---|---|---|
| **(4, 2) - groups** | 48 | 6 | 0 | 4 | 1 | 6 | 5 | 11 | 3 |
| **(3, 2) – groups; real order** | 96 | 6 | 1 | 10 | 0 | 7 | 10 | 17 | 2 |
| **(3, 2) – groups; 1-2-3 order** | 96 | 9 | 0 | 1 | 1 | 9 | 2 | 11 | 2 |
| **(3, 2) – groups; real order; red groups** | 51 | 5 | 1 | 4 | 0 | 6 | 4 | 10 | 2 |
| **(3, 2) – groups; 1-2-3 order; red groups** | 51 | 7 | 0 | 0 | 1 | 7 | 1 | 8 | 2 |

**Fig. A.9.** 2018 FIFA World Cup - Non-competitive matches summary.

| 2018 World Cup | *C* | *WC* | *D* | *WD* | *TC* | *TD* | *TC+TD* | *MTC* |
|---|---|---|---|---|---|---|---|---|
| **(4, 2) - groups** | 6 | 0 | 4 | 1 | 6 | 5 | 11 | 3 |
| **(3, 2) – groups; real order** | 3 | 0.5 | 5 | 0 | 3.5 | 5 | 8.5 | 1 |
| **(3, 2) – groups; 1-2-3 order** | 4.5 | 0 | 0.5 | 0.5 | 4.5 | 1 | 5.5 | 1 |
| **(3, 2) – groups; real order; red groups** | 4.71 | 0.94 | 3.76 | 0.00 | 5.65 | 3.76 | 9.41 | 1.88 |
| **(3, 2) – groups; 1-2-3 order; red groups** | 6.59 | 0.00 | 0.00 | 0.94 | 6.59 | 0.94 | 7.53 | 1.88 |

**Fig. A.10.** 2018 FIFA World Cup - Non-competitive matches summary weighted to 48 matches (2026 World Cup format).

# Appendix B – 2014 FIFA World Cup per Group Analysis for Non-Competitive Matches

In Appendix B, we show the non-competitive matches that could occur in all possible (3, 2)-groups that are based on actual (4, 2)-groups in the 2014 FIFA World Cup [7]. Information regarding the non-competitive match is included in below images. Empty cells means there were no matches in that category. The matches' summaries are shown in the last two figures.

| Group | C | WC | D | WD |
|---|---|---|---|---|
| A | BRA – CMR (BRA win/draw), **yes** | | | |
| A1a | | | | |
| A2a | | | | BRA – CMR, CMR 3+ win |
| A3a | | | | |
| A4a | | | MEX – CMR | |
| A1b | | | | |
| A2b | | | | |
| A3b | | | | |
| A4b | | | | MEX – CMR, CMR 3+ win |

**Fig. B.1.** 2014 FIFA World Cup - Group **A** analysis of non-competitive matches.

| Group | C | WC | D | WD |
|---|---|---|---|---|
| B | | | ESP – AUS<br>NED – CHI | |
| B1a | | | NED – CHI | |
| B2a | | | | |
| B3a | | | | |
| B4a | | | NED – CHI | |
| B1b | | | NED – CHI | |
| B2b | NED – AUS (0-3), no | | | |
| B3b | | | | |
| B4b | | | | |

**Fig. B.2.** 2014 FIFA World Cup - Group **B** analysis of non-competitive matches.

| Group | C | WC | D | WD |
|---|---|---|---|---|
| C | COL – JPN (JPN win), no | | | |
| C1a | | | | |
| C2a | COL – JPN (most), **yes (4-1)** | | | |
| C3a | COL – JPN (2-3), no | | | |
| C4a | GRE – CIV (most), **yes (2-1)** | | | |
| C1b | | | | |
| C2b | | | | |
| C3b | | | | |
| C4b | | | | |

**Fig. B.3.** 2014 FIFA World Cup - Group **C** analysis of non-competitive matches.

| Group | C | WC | D | WD |
|---|---|---|---|---|
| **D** | | | ENG - CRC | |
| **D1a** | | | URU – ITA | |
| **D2a** | | | | |
| **D3a** | ENG – CRC (1-0), no | | | |
| **D4a** | ENG – CRC (2-1), no | | | |
| **D1b** | | | | |
| **D2b** | ITA – CRC (1-0), no | | | |
| **D3b** | ENG – CRC (1-0), no | | | |
| **D4b** | ENG – CRC (2-1), no | | | |

**Fig. B.4.** 2014 FIFA World Cup - Group **D** analysis of non-competitive matches.

| Group | C | WC | D | WD |
|---|---|---|---|---|
| **E** | | | | |
| **E1a** | FRA – ECU (0-1), no | | | |
| **E2a** | | | | |
| **E3a** | SUI – HON (2 – 3) , no | | | |
| E4a | | | 1 (F – E) | |
| **E1b** | FRA – ECU (0-1), no | | | |
| **E2b** | FRA – HON (3 - 6), no | | | |
| **E3b** | | | | |
| **E4b** | | | | |

**Fig. B.5.** 2014 FIFA World Cup - Group **E** analysis of non-competitive matches.

| Group | C | WC | D | WD |
|---|---|---|---|---|
| **F** | ARG – NGR (NGR win), no | | | BIH – IRN |
| **F1a** | | | ARG – NGR | |
| **F2a** | | | | |
| **F3a** | ARG – NGR (most), **yes (3-2)** | | | |
| **F4a** | | | | |
| **F1b** | | | | |
| **F2b** | | | | |
| **F3b** | | | | |
| **F4b** | | | | |

**Fig. B.6.** 2014 FIFA World Cup - Group **F** analysis of non-competitive matches.

| Group | C | WC | D | WD |
|---|---|---|---|---|
| **G** | | | | |
| **G1a** | USA – GHA (most), **yes (2-1)** | | | |
| **G2a** | | | | |
| **G3a** | USA – GHA (most), **yes (2-1)** | | | |
| **G4a** | | | | |
| **G1b** | | | | |
| **G2b** | | | | |
| **G3b** | | | | |
| **G4b** | | | | |

**Fig. B.7.** 2014 FIFA World Cup - Group **G** analysis of non-competitive matches.

| Group | C | WC | D | WD |
|---|---|---|---|---|
| H | BEL – KOR (KOR win), no | | | |
| H1a | | | | |
| H2a | BEL – KOR (1-2), no | | | |
| H3a | | | | |
| H4a | RUS – ALG (most), yes (1-1) | | | |
| H1b | | | | |
| H2b | | | | |
| H3b | | | | |
| H4b | ALG – KOR (2 – 2), no | | | |

**Fig. B.8.** 2014 FIFA World Cup - Group **H** analysis of non-competitive matches.

| 2014 World Cup | Total Matches | C | WC | D | WD | TC | TD | TC+TD | MTC |
|---|---|---|---|---|---|---|---|---|---|
| (4, 2) - groups | 48 | 4 | 0 | 3 | 1 | 4 | 4 | 8 | 1 |
| (3, 2) – groups; real order | 96 | 13 | 0 | 5 | 1 | 13 | 6 | 19 | 6 |
| (3, 2) – groups; 1-2-3 order | 96 | 6 | 0 | 2 | 1 | 6 | 3 | 9 | 0 |
| (3, 2) – groups; real order; red groups | 39 | 6 | 0 | 2 | 0 | 6 | 2 | 8 | 2 |
| (3, 2) – groups; 1-2-3 order; red groups | 39 | 2 | 0 | 0 | 1 | 2 | 1 | 3 | 0 |

**Fig. B.9.** 2014 FIFA World Cup - Non-competitive matches summary.

| 2014 World Cup | C | WC | D | WD | TC | TD | TC+TD | MTC |
|---|---|---|---|---|---|---|---|---|
| (4, 2) - groups | 4 | 0 | 3 | 1 | 4 | 4 | 8 | 1 |
| (3, 2) – groups; real order | 6.5 | 0 | 2.5 | 0.5 | 6.5 | 3 | 9.5 | 3 |
| (3, 2) – groups; 1-2-3 order | 3 | 0 | 1 | 0.5 | 3 | 1.5 | 4.5 | 0 |
| (3, 2) – groups; real order; red groups | 4.59 | 0.00 | 1.53 | 0.00 | 4.59 | 1.53 | 6.12 | 1.53 |
| (3, 2) – groups; 1-2-3 order; red groups | 1.53 | 0.00 | 0.00 | 0.76 | 1.53 | 0.76 | 2.29 | 0.00 |

**Fig. B.10.** 2014 FIFA World Cup - Non-competitive matches summary weighted to 48 matches (2026 FIFA World Cup format).

# Appendix C –2010 FIFA World Cup per Group Analysis for Non-Competitive Matches

In Appendix C, we show the non-competitive matches that could occur in all possible (3, 2)-groups that are based on actual (4, 2)-groups in the 2010 FIFA World Cup [7]. Information regarding the non-competitive match is included in below images. Empty cells means there were no matches in that category. The matches' summaries are shown in the last two figures.

| Group | *C* | *WC* | *D* | *WD* |
|---|---|---|---|---|
| **A** | MEX – URU (draw), no | | | FRA – RSA |
| **A1a** | | | | |
| **A2a** | | | | |
| **A3a** | MEX – URU (1-0), no | | | |
| **A4a** | MEX – URU (0-1), **yes** | | | |
| **A1b** | | | | |
| **A2b** | FRA – URU (1-0), no | | | |
| **A3b** | MEX – URU (1-0), no | | | |
| **A4b** | MEX – URU (0-1), **yes** | | | |

**Fig. C.1.** 2010 FIFA World Cup - Group **A** analysis of non-competitive matches.

| Group | *C* | *WC* | *D* | *WD* |
|---|---|---|---|---|
| **B** | | | | |
| **B1a** | | | ARG – GRE | |
| **B2a** | | | | |
| **B3a** | | | | |
| **B4a** | NGR – KOR (1-0), no | | | |
| **B1b** | | | | |
| **B2b** | | | | |
| **B3b** | | | | |
| **B4b** | NGR – KOR (1-0), no | | | |

**Fig. C.2.** 2010 FIFA World Cup - Group **B** analysis of non-competitive matches.

| Group | *C* | *WC* | *D* | *WD* |
|---|---|---|---|---|
| **C** | | | | |
| **C1a** | USA – ALG (2-2), no | | | |
| **C2a** | ENG – SLO (3-3), no | | | |
| **C3a** | ENG – SLO (most), **yes (1-0)** | | | |
| **C4a** | | | | |
| **C1b** | USA – ALG (2-2), no | | | |
| **C2b** | ENG – SLO (3-3), no | | | |
| **C3b** | | | | |
| **C4b** | | | | |

**Fig. C.3.** 2010 FIFA World Cup - Group **C** analysis of non-competitive matches.

| Group | *C* | *WC* | *D* | *WD* |
|---|---|---|---|---|
| **D** | | | | |
| **D1a** | | | | SRB - AUS, AUS +3 win |
| **D2a** | GER –GHA (1-0), **yes*** | | | |
| **D3a** | GER –GHA (most), **yes (1-0)** | | | |
| **D4a** | | | | |
| **D1b** | | | | SRB - AUS, AUS +3 win |
| **D2b** | SRB – GHA (1-2), no* | | | |
| **D3b** | | | | |
| **D4b** | | | AUS - GHA | |

*\* Ranking determined based on disciplinary record*

**Fig. C.4.** 2010 FIFA World Cup - Group **D** analysis of non-competitive matches.

| Group | *C* | *WC* | *D* | *WD* |
|---|---|---|---|---|
| **E** | | | NED – CMR | |
| **E1a** | NED – CMR (0 – 1), no | | | |
| **E2a** | | | | |
| **E3a** | NED – CMR (1 – 2), no | | | |
| **E4a** | | | DEN - JPN | |
| **E1b** | | | | |
| **E2b** | | | | |
| **E3b** | | | | |
| **E4b** | | | DEN - JPN | |

**Fig. C.5.** 2010 FIFA World Cup - Group **E** analysis of non-competitive matches.

| Group | *C* | *WC* | *D* | *WD* |
|---|---|---|---|---|
| **F** | | | | |
| **F1a** | | | | |
| **F2a** | PAR-NZL (2-2), no | | | |
| **F3a** | ITA –SVK (2-2), no | | | |
| **F4a** | PAR-NZL (many), **yes (0-0)** | | | |
| **F1b** | PAR-SVK (many), **yes (2-0)** | | | |
| **F2b** | | | | |
| **F3b** | PAR-SVK (many), **yes (2-0)** | | | |
| **F4b** | | | | |

**Fig. C.6.** 2010 FIFA World Cup - Group **F** analysis of non-competitive matches.

| Group | *C* | *WC* | *D* | *WD* |
|---|---|---|---|---|
| **G** | BRA – POR (POR win/draw), **yes (0-0)** | | | |
| **G1a** | BRA – POR (most), **yes (0-0)** | | | |
| **G2a** | | | BRA – POR | |
| **G3a** | | | | |
| **G4a** | | | | |
| **G1b** | | | | |
| **G2b** | | | | |
| **G3b** | | | | |
| **G4b** | | | | |

**Fig. C.7.** 2010 FIFA World Cup - Group **G** analysis of non-competitive matches.

| Group | C | WC | D | WD |
|---|---|---|---|---|
| **H** | | | | |
| **H1a** | ESP – CHI (2 – 1), **yes** | | | |
| **H2a** | | | ESP – CHI | |
| **H3a** | | | | |
| **H4a** | | | | |
| **H1b** | CHI – SUI (3 – 2), no | | | |
| **H2b** | | | | |
| **H3b** | | | | |
| **H4b** | | | | |

**Fig. C.8.** 2010 FIFA World Cup - Group **H** analysis of non-competitive matches.

| 2010 World Cup | Total Matches | C | WC | D | WD | TC | TD | TC+TD | MTC |
|---|---|---|---|---|---|---|---|---|---|
| **(4, 2) - groups** | 48 | 2 | 0 | 1 | 2 | 2 | 3 | 5 | 1 |
| **(3, 2) – groups; real order** | 96 | 15 | 0 | 4 | 1 | 15 | 5 | 20 | 7 |
| **(3, 2) – groups; 1-2-3 order** | 96 | 9 | 0 | 2 | 1 | 9 | 3 | 12 | 3 |
| **(3, 2) – groups; real order; red groups** | 51 | 8 | 0 | 2 | 1 | 8 | 3 | 11 | 4 |
| **(3, 2) – groups; 1-2-3 order; red groups** | 51 | 5 | 0 | 1 | 1 | 5 | 2 | 7 | 2 |

**Fig. C.9.** 2010 FIFA World Cup - Non-competitive matches summary.

| 2010 World Cup | C | WC | D | WD | TC | TD | TC+TD | MTC |
|---|---|---|---|---|---|---|---|---|
| **(4, 2) - groups** | 2 | 0 | 1 | 2 | 2 | 3 | 5 | 1 |
| **(3, 2) – groups; real order** | 7.5 | 0 | 2 | 0.5 | 7.5 | 2.5 | 10 | 3.5 |
| **(3, 2) – groups; 1-2-3 order** | 4.5 | 0 | 1 | 0.5 | 4.5 | 1.5 | 6 | 1.5 |
| **(3, 2) – groups; real order; red groups** | 7.53 | 0.00 | 1.88 | 0.94 | 7.53 | 2.82 | 10.35 | 3.76 |
| **(3, 2) – groups; 1-2-3 order; red groups** | 4.71 | 0.00 | 0.94 | 0.94 | 4.71 | 1.88 | 6.59 | 1.88 |

**Fig. C.10.** 2010 FIFA World Cup - Non-competitive matches summary weighted to 48 matches (2026 FIFA World Cup format).

# Appendix D –2006 FIFA World Cup per Group Analysis for Non-Competitive Matches

In Appendix D, we show the non-competitive matches that could occur in all possible (3, 2)-groups that are based on actual (4, 2)-groups in the 2006 FIFA World Cup [7]. Information regarding the non-competitive match is included in below images. Empty cells means there were no matches in that category. The matches' summaries are shown in the last two figures.

| Group | C | WC | D | WD |
|---|---|---|---|---|
| A | | | GER – ECU<br>CRC - POL | |
| A1a | | | | |
| A2a | | | GER – ECU | |
| A3a | | | GER – ECU | |
| A4a | | | | |
| A1b | | | | |
| A2b | | | | |
| A3b | | | | |
| A4b | | | POL - ECU | |

**Fig. D.1.** 2006 FIFA World Cup - Group **A** analysis of non-competitive matches.

| Group | C | WC | D | WD |
|---|---|---|---|---|
| B | ENG – SWE (SWE win/draw), yes (2-2) | | | PAR - TRI |
| B1a | | | ENG - SWE | |
| B2a | ENG – SWE (SWE win/draw), yes (2-2) | | | |
| B3a | | | | |
| B4a | | | | |
| B1b | | | | |
| B2b | | | | |
| B3b | | | | |
| B4b | | | | |

**Fig. D.2.** 2006 FIFA World Cup - Group **B** analysis of non-competitive matches.

| Group | C | WC | D | WD |
|---|---|---|---|---|
| C | | | NED – ARG<br>CIV – SCG | |
| C1a | | | NED – ARG | |
| C2a | | | NED – ARG | |
| C3a | | | | |
| C4a | | | | |
| C1b | | | | |
| C2b | | | | |
| C3b | | | | |
| C4b | | | | |

**Fig. D.3.** 2006 FIFA World Cup - Group **C** analysis of non-competitive matches.

| Group | C | WC | D | WD |
|---|---|---|---|---|
| D | MEX – POR (MEX win/draw), no | | | IRN - ANG |
| D1a | | | MEX – POR | |
| D2a | MEX – POR (most), yes (1-2) | | | |
| D3a | | | | |
| D4a | | | | |
| D1b | | | | |
| D2b | POR – ANG (ANG win/draw), no | | | |
| D3b | | | | |
| D4b | | | | |

Fig. **D.4.** 2006 FIFA World Cup - Group **D** analysis of non-competitive matches.

| Group | C | WC | D | WD |
|---|---|---|---|---|
| E | CZE – ITA (ITA win/draw), yes (0-2) | | | |
| E1a | | | | |
| E2a | | | USA - GHA | |
| E3a | CZE – ITA (3-1), no | | | |
| E4a | | | | |
| E1b | | | USA – ITA | |
| E2b | | | USA - GHA | |
| E3b | | | ITA - GHA | |
| E4b | ITA – GHA (ITA win/draw), yes (2-0) | | | |

Fig. **D.5.** 2006 FIFA World Cup - Group **E** analysis of non-competitive matches.

| Group | C | WC | D | WD |
|---|---|---|---|---|
| F | BRA – JPN (JPN win), no | | | |
| F1a | BRA – JPN (JPN win/draw), no | | | |
| F2a | BRA – JPN (2-4), no | | | |
| F3a | | | | |
| F4a | CRO – AUS (CRO win/draw), yes (2-2) | | | |
| F1b | | | | |
| F2b | | | | |
| F3b | | | | |
| F4b | CRO – AUS (CRO win/draw), yes (2-2) | | | |

Fig. **D.6.** 2006 FIFA World Cup - Group **F** analysis of non-competitive matches.

| Group | C | WC | D | WD |
|---|---|---|---|---|
| G | FRA – TOG (FRA win by 2), yes (2-0) | | | |
| G1a | KOR – SUI (2-2), no | | | |
| G2a | | | | |
| G3a | | | | |
| G4a | | | KOR – SUI | |
| G1b | KOR – SUI (2-2), no | | | |
| G2b | | | | |
| G3b | | | | |
| G4b | SUI – TOG (0-1), no | | | |

Fig. **D.7.** 2006 FIFA World Cup - Group **G** analysis of non-competitive matches.

| Group | C | WC | D | WD |
|---|---|---|---|---|
| H | ESP – KSA (KSA win), no | | | TUN - UKR |
| H1a | ESP – KSA (most), yes (1-0) | | | |
| H2a | | ESP – TUN (0-1), no | | |
| H3a | ESP – KSA (1-5), no | | | |
| H4a | TUN – UKR (most), yes (0-1) | | | |
| H1b | | | | |
| H2b | | | | |
| H3b | | | | |
| H4b | KSA – UKR (KSA win/draw), no | | | |

**Fig. D.8.** 2006 FIFA World Cup - Group **H** analysis of non-competitive matches.

| 2006 World Cup | Total Matches | C | WC | D | WD | TC | TD | TC+TD | MTC |
|---|---|---|---|---|---|---|---|---|---|
| (4, 2) - groups | 48 | 6 | 0 | 4 | 3 | 6 | 7 | 13 | 3 |
| (3, 2) – groups; real order | 96 | 10 | 1 | 8 | 0 | 11 | 8 | 19 | 5 |
| (3, 2) – groups; 1-2-3 order | 96 | 6 | 0 | 4 | 0 | 6 | 4 | 10 | 2 |
| (3, 2) – groups; real order; red groups | 51 | 4 | 1 | 2 | 0 | 5 | 2 | 7 | 2 |
| (3, 2) – groups; 1-2-3 order; red groups | 51 | 2 | 0 | 0 | 0 | 2 | 0 | 2 | 1 |

**Fig. D.9.** 2006 FIFA World Cup - Non-competitive matches summary.

| 2006 World Cup | C | WC | D | WD | TC | TD | TC+TD | MTC |
|---|---|---|---|---|---|---|---|---|
| (4, 2) - groups | 6 | 0 | 4 | 3 | 6 | 7 | 13 | 3 |
| (3, 2) – groups; real order | 5 | 0.5 | 4 | 0 | 5.5 | 4 | 9.5 | 2.5 |
| (3, 2) – groups; 1-2-3 order | 3 | 0 | 2 | 0 | 3 | 2 | 5 | 1 |
| (3, 2) – groups; real order; red groups | 3.76 | 0.94 | 1.88 | 0.00 | 4.71 | 1.88 | 6.59 | 1.88 |
| (3, 2) – groups; 1-2-3 order; red groups | 1.88 | 0.00 | 0.00 | 0.00 | 1.88 | 0.00 | 1.88 | 0.94 |

**Fig. D.10.** 2006 FIFA World Cup - Non-competitive matches summary weighted to 48 matches (2026 FIFA World Cup format).

# Appendix E –2002 FIFA World Cup per Group Analysis for Non-Competitive Matches

In Appendix E, we show the non-competitive matches that could occur in all possible (3, 2)-groups that are based on actual (4, 2)-groups in the 2002 FIFA World Cup [7]. Information regarding the non-competitive match is included in below images. Empty cells means there were no matches in that category. The matches' summaries are shown in the last two figures.

| Group | C | WC | D | WD |
|---|---|---|---|---|
| **A** | | | | |
| **A1a** | FRA – DEN (FRA win/draw), no | | | |
| **A2a** | | | | |
| **A3a** | URU – SEN (URU win/draw), yes (3-3) | | | |
| **A4a** | | | | |
| **A1b** | DEN – URU (most), yes (2-1) | | | |
| **A2b** | | | DEN - SEN | |
| **A3b** | URU – SEN (URU win/draw), yes (3-3) | | | |
| **A4b** | | | | |

**Fig. E.1.** 2002 FIFA World Cup - Group **A** analysis of non-competitive matches.

| Group | C | WC | D | WD |
|---|---|---|---|---|
| **B** | ESP – RSA (RSA win/draw), no | | | PAR - SLO |
| **B1a** | | | | |
| **B2a** | ESP – RSA (most), yes (3-2) | | | |
| **B3a** | ESP – RSA (most), yes (3-2) | | | |
| **B4a** | | | | |
| **B1b** | | | | |
| **B2b** | | | | |
| **B3b** | | | | |
| **B4b** | | | | |

**Fig. E.2.** 2002 FIFA World Cup - Group **B** analysis of non-competitive matches.

| Group | C | WC | D | WD |
|---|---|---|---|---|
| **C** | BRA – CRC (CRC win/draw), no | | | TUR - CHN |
| **C1a** | BRA – CRC (CRC win/draw), no | | | |
| **C2a** | | | | |
| **C3a** | | | BRA - CRC | |
| **C4a** | | | | |
| **C1b** | | | | |
| **C2b** | | | | |
| **C3b** | | | | |
| **C4b** | | | | |

**Fig. E.3.** 2002 FIFA World Cup - Group **C** analysis of non-competitive matches.

| Group | C | WC | D | WD |
|---|---|---|---|---|
| **D** | USA – POL (USA win/draw), no | | | |
| **D1a** | | | | |
| **D2a** | | | KOR - POR | |
| **D3a** | | | | |
| **D4a** | | | | USA – POL. POL 3+ win |
| **D1b** | | | | |
| **D2b** | | | | |
| **D3b** | | | | |
| **D4b** | | | | USA – POL. POL 3+ win |

**Fig. E.4.** 2002 FIFA World Cup - Group **D** analysis of non-competitive matches.

| Group | C | WC | D | WD |
|---|---|---|---|---|
| **E** | | | | |
| **E1a** | | GER – IRL (IRL win), no | | |
| **E2a** | GER – CMR (2-2), no | | | |
| **E3a** | | | GER - CMR | |
| **E4a** | | | | |
| **E1b** | | | | |
| **E2b** | | | | |
| **E3b** | | GER – CMR (CMR win), no | | |
| **E4b** | | | | |

**Fig. E.5.** 2002 FIFA World Cup - Group **E** analysis of non-competitive matches.

| Group | C | WC | D | WD |
|---|---|---|---|---|
| **F** | ENG – NGA (ENG win/draw), yes (0-0) | | | |
| **F1a** | | | | |
| **F2a** | ENG – NGA (1-2), no | | | |
| **F3a** | | | ARG - SWE | |
| **F4a** | | | | |
| **F1b** | ENG – SWE (SWE win/draw), yes (1-1) | | | |
| **F2b** | ENG – NGA (1-2), no | | | |
| **F3b** | | | | |
| **F4b** | SWE – NGA (2-2), no | | | |

**Fig. E.6.** 2002 FIFA World Cup - Group **F** analysis of non-competitive matches.

| Group | C | WC | D | WD |
|---|---|---|---|---|
| **G** | | | | |
| **G1a** | ITA – MEX (3-2), no | | | |
| **G2a** | | | ITA - MEX | |
| **G3a** | | | | |
| **G4a** | | | | |
| **G1b** | MEX – CRO (MEX win/draw), yes (1-0) | | | |
| **G2b** | | | | |
| **G3b** | | | | |
| **G4b** | | | | |

**Fig. E.7.** 2002 FIFA World Cup - Group **G** analysis of non-competitive matches.

| Group | C | WC | D | WD |
|---|---|---|---|---|
| **H** | | | | |
| **H1a** | | | | |
| **H2a** | JPN – TUN (3-3), no | | | |
| **H3a** | | | | |
| **H4a** | BEL – RUS (BEL win/draw), yes (3-2) | | | |
| **H1b** | | | | |
| **H2b** | | | | |
| **H3b** | | | | |
| **H4b** | | | | |

**Fig. E.8.** 2002 FIFA World Cup - Group **H** analysis of non-competitive matches.

| 2002 World Cup | Total Matches | C | WC | D | WD | TC | TD | TC+TD | MTC |
|---|---|---|---|---|---|---|---|---|---|
| **(4, 2) - groups** | 48 | 4 | 0 | 0 | 2 | 4 | 2 | 6 | 1 |
| **(3, 2) – groups; real order** | 96 | 10 | 1 | 5 | 1 | 11 | 6 | 17 | 4 |
| **(3, 2) – groups; 1-2-3 order** | 96 | 6 | 1 | 1 | 1 | 7 | 2 | 9 | 4 |
| **(3, 2) – groups; real order; red groups** | 60 | 6 | 1 | 3 | 1 | 7 | 4 | 11 | 2 |
| **(3, 2) – groups; 1-2-3 order; red groups** | 60 | 3 | 1 | 1 | 1 | 4 | 2 | 6 | 2 |

**Fig. E.9.** 2002 FIFA World Cup - Non-competitive matches summary.

| 2002 World Cup | C | WC | D | WD | TC | TD | TC+TD | MTC |
|---|---|---|---|---|---|---|---|---|
| **(4, 2) - groups** | 4 | 0 | 0 | 2 | 4 | 2 | 6 | 1 |
| **(3, 2) – groups; real order** | 5 | 0.5 | 2.5 | 0.5 | 5.5 | 3 | 8.5 | 2 |
| **(3, 2) – groups; 1-2-3 order** | 3 | 0.5 | 0.5 | 0.5 | 3.5 | 1 | 4.5 | 2 |
| **(3, 2) – groups; real order; red groups** | 4.80 | 0.80 | 2.40 | 0.80 | 5.60 | 3.20 | 8.80 | 1.60 |
| **(3, 2) – groups; 1-2-3 order; red groups** | 2.40 | 0.80 | 0.80 | 0.80 | 3.20 | 1.60 | 4.80 | 1.60 |

**Fig. E.10.** 2002 FIFA World Cup - Non-competitive matches summary weighted to 48 matches (2026 FIFA World Cup format).

# Appendix F –1998 FIFA World Cup per Group Analysis for Non-Competitive Matches

In Appendix F, we show the non-competitive matches that could occur in all possible (3, 2)-groups that are based on actual (4, 2)-groups in the 1998 FIFA World Cup [7]. Information regarding the non-competitive match is included in below images. Empty cells means there were no matches in that category. The matches' summaries are shown in the last two figures.

| Group | C | WC | D | WD |
|---|---|---|---|---|
| A | BRA – NOR (NOR win), yes (1-2)* | | | MAR – SCO |
| A1a | BRA – NOR (most), yes (1-2) | | | |
| A2a | BRA – NOR (Nor win, draw), yes (1-2) | | | |
| A3a | | | | |
| A4a | MAR – SCO (3-3), no | | | |
| A1b | | | | |
| A2b | NOR – SCO (2-3), no | | | |
| A3b | | | | |
| A4b | MAR – SCO (3-3), no | | | |

*Aggravated collusion*

**Fig. F.1.** 1998 FIFA World Cup - Group **A** analysis of non-competitive matches.

| Group | C | WC | D | WD |
|---|---|---|---|---|
| B | | | | |
| B1a | ITA – AUT (3-3), no | | | |
| B2a | | | | |
| B3a | ITA – AUT (most), yes (2-1) | | | |
| B4a | CHI – CMR (2-2), no | | | |
| B1b | | | | |
| B2b | | | | |
| B3b | | | | |
| B4b | AUT – CMR (1-1), yes* | | | |

*Ranking determined based on disciplinary record*

**Fig. F.2.** 1998 FIFA World Cup - Group **B** analysis of non-competitive matches.

| Group | C | WC | D | WD |
|---|---|---|---|---|
| C | FRA – DEN (DEN win/draw), no | | | KSA – RSA |
| C1a | | | FRA – DEN | |
| C2a | FRA – DEN (most), yes (2-1) | | | |
| C3a | | | | |
| C4a | | | | |
| C1b | | | | |
| C2b | | | | |
| C3b | | | | |
| C4b | | | | |

**Fig. F.3.** 1998 FIFA World Cup - Group **C** analysis of non-competitive matches.

| Group | C | WC | D | WD |
|---|---|---|---|---|
| **D** | PAR – NGA (PAR win), yes (3-1)* | | | ESP – BUL |
| **D1a** | ESP – BUL (1-1), no | | | |
| **D2a** | | | | |
| **D3a** | | | | |
| **D4a** | PAR – NGA (PAR win/draw), yes (3-1) | | | |
| **D1b** | | | | |
| **D2b** | | ESP – NGA (NGA win), yes (2-3) | | BUL – NGA, BUL 3+ win |
| **D3b** | | | | |
| **D4b** | PAR – NGA (PAR win/draw), yes (3-1) | | | |

*Aggravated collusion*

**Fig. F.4.** 1998 FIFA World Cup - Group **D** analysis of non-competitive matches.

| Group | C | WC | D | WD |
|---|---|---|---|---|
| **E** | | MEX – NED (draw), yes (2-2) | | KOR – BEL, BEL 3+ win |
| **E1a** | MEX - NED | | | |
| **E2a** | MEX – NED (2-2), yes* | | | |
| **E3a** | | | | |
| **E4a** | | | | |
| **E1b** | | | | |
| **E2b** | NED - BEL (3-3), no | | | |
| **E3b** | | | | |
| **E4b** | | NED – BEL (BEL win), no | | |

**Fig. F.5.** 1998 FIFA World Cup - Group **E** analysis of non-competitive matches.

| Group | C | WC | D | WD |
|---|---|---|---|---|
| **F** | YUG – USA (YUG win/draw), yes (1-0) | | | |
| **F1a** | | | | |
| **F2a** | | | | |
| **F3a** | | | GER - IRN | |
| **F4a** | YUG-USA (2-3), no | | | |
| **F1b** | | | | |
| **F2b** | | | | |
| **F3b** | | | | |
| **F4b** | | | | |

**Fig. F.6.** 1998 FIFA World Cup - Group **F** analysis of non-competitive matches.

| Group | C | WC | D | WD |
|---|---|---|---|---|
| **G** | | | ROU – TUN | |
| **G1a** | | | | |
| **G2a** | | | | |
| **G3a** | ROU – TUN (1-2), no | | | |
| **G4a** | | | ENG - COL | |
| **G1b** | | | | |
| **G2b** | | | | |
| **G3b** | | | | |
| **G4b** | | | | |

**Fig. F.7.** 1998 FIFA World Cup - Group **G** analysis of non-competitive matches.

| Group | C | WC | D | WD |
|---|---|---|---|---|
| H | | | ARG – CRO<br>JPN – JAM | |
| H1a | | | ARG – CRO | |
| H2a | | | | |
| H3a | | | | |
| H4a | | | ARG – CRO | |
| H1b | | | ARG – CRO | |
| H2b | | | ARG – JAM | |
| H3b | | | CRO – JAM | |
| H4b | | | | |

**Fig. F.8.** 1998 FIFA World Cup - Group **H** analysis of non-competitive matches.

| 1998 World Cup | Total Matches | C | WC | D | WD | TC | TD | TC+TD | MTC |
|---|---|---|---|---|---|---|---|---|---|
| (4, 2) - groups | 48 | 4 | 1 | 3 | 4 | 5 | 7 | 12 | 4 |
| (3, 2) – groups; real order | 96 | 13 | 0 | 5 | 0 | 13 | 5 | 18 | 6 |
| (3, 2) – groups; 1-2-3 order | 96 | 4 | 2 | 4 | 0 | 6 | 4 | 10 | 2 |
| (3, 2) – groups; real order; red groups | 39 | 3 | 0 | 1 | 0 | 3 | 1 | 4 | 1 |
| (3, 2) – groups; 1-2-3 order; red groups | 39 | 0 | 2 | 0 | 1 | 2 | 1 | 3 | 1 |

**Fig. F.9.** 1998 FIFA World Cup - Non-competitive matches summary.

| 1998 World Cup | C | WC | D | WD | TC | TD | TC+TD | MTC |
|---|---|---|---|---|---|---|---|---|
| (4, 2) - groups | 4 | 1 | 3 | 4 | 5 | 7 | 12 | 4 |
| (3, 2) – groups; real order | 6.5 | 0 | 2.5 | 0 | 6.5 | 2.5 | 9 | 3 |
| (3, 2) – groups; 1-2-3 order | 2 | 1 | 2 | 0 | 3 | 2 | 5 | 1 |
| (3, 2) – groups; real order; red groups | 3.69 | 0.00 | 1.23 | 0.00 | 3.69 | 1.23 | 4.92 | 1.23 |
| (3, 2) – groups; 1-2-3 order; red groups | 0.00 | 2.46 | 0.00 | 1.23 | 2.46 | 1.23 | 3.69 | 1.23 |

**Fig. F.10.** 1998 FIFA World Cup - Non-competitive matches summary weighted to 48 matches (2026 FIFA World Cup format).